\documentclass[sigconf,nonacm]{acmart}

\usepackage{graphicx}
\usepackage[disable]{todonotes}
\usepackage{listings}
\usepackage{multirow}
\usepackage{amsmath}
\usepackage{stmaryrd}
\usepackage{mdframed}
\usepackage{tcolorbox}
\usepackage{multicol}
\usepackage{caption}
\usepackage{xspace}
\usepackage{url}
\usepackage{amsthm}
\usepackage{dirtree}
\usepackage[bottom]{footmisc}
\usepackage[inline]{enumitem}

\lstdefinestyle{mystyle}{
	basicstyle=\ttfamily\scriptsize,
	breakatwhitespace=false,
	breaklines=true,
	captionpos=b,
	keepspaces=true,
	numbers=left,
	numbersep=5pt,
	showspaces=false,
	showstringspaces=false,
	showtabs=false,
	tabsize=2,
	frame=single,
	xleftmargin=3.0ex
}

\newcommand{\bh}[1]{}
\newcommand{\jd}[1]{}

\newcommand{\bfs}{\textit{obfs}\xspace}
\newcommand{\gaoss}{\textit{gaoss}\xspace}
\newcommand{\mvnc}{\textit{mvnc}\xspace}

\title{On the Variability of Source Code in Maven Package Rebuilds}

\author{Jens Dietrich}
\email{jens.dietrich@vuw.ac.nz}
\affiliation{%
	\institution{Victoria University of Wellington}
	\city{Wellington}
	\country{New Zealand}
}
\author{Behnaz Hassanshahi}
\email{behnaz.hassanshahi@oracle.com}
\affiliation{%
	\institution{Oracle Inc}
	\city{Brisbane}
	\country{Australia}
}

\begin{abstract}
	Rebuilding packages from open source is a common practice to improve the security of software supply chains, and is now done at an industrial scale. The basic principle is to acquire the source code used to build a package published in a repository such as Maven Central (for Java), rebuild the package independently with hardened security, and publish it in some alternative repository. In this paper we test the assumption that the same source code is being used by those alternative builds.
	To study this, we compare the sources released with packages on Maven Central, with the sources associated with independently built packages from Google's Assured Open Source and Oracle's Build-from-Source projects.
	We study non-equivalent sources for alternative builds of 28 popular packages with 85 releases.
	We investigate the causes of non-equivalence, and find that the main cause is build extensions that generate code at build time, which are difficult to reproduce.
	We suggest strategies to address this issue.
\end{abstract}


\begin{document}
\maketitle

\section{Introduction}
\label{sec:introduction}

Modern software engineering heavily relies on open-source packages and automated processes to build and distribute applications. 
While this enables economies of scale and rapid innovation, it also introduces significant security risks.
Incidents and vulnerabilities like  \textit{solarwinds}, \textit{codecov}, \textit{equifax}, \textit{log4shell}~\cite{ellison2010evaluating,martinez2021software,enck2022top} and more recently \textit{react2shell}~\cite{react2shell} have raised major concerns about the security of software supply chains~\cite{EO14028}. This includes the security of 
complex and heavily automated processes used to build and deploy software. 
The possibility of such attacks has been hypothesised early in Ken Thompson's famous Turning award speech~\cite{thompson1984reflections}. Such attacks have now become common, examples include \textit{xcodeghost}, \textit{ccleaner}, \textit{shadowpad}, \textit{shai-hulud}, \textit{shadowhammer} and \textit{xz}~\cite{martinez2021software,andreoli2023prevalence,xz,ohm2020backstabber}.

A possible approach to detect packages produced by compromised builds is to reproduce builds~\cite{reproduciblebuilds,lamb2021reproducible}.
I.e. a build is repeated, and the packages produced by the original and the secondary build are compared, usually by means of cryptographic hashes. 
If an attacker had compromised a build, then a second (non-compromised) build would produce a different binary, and this can be easily detected~\cite{enck2022top, butler2023business, fourne2023s}. 
The following rule expresses the basic idea: two builds starting from the same sources should produce the same binaries.

\vspace{-0.3cm}
\begin{equation}
	src_1 = src_2 \Rightarrow build_1(src_1) = build_2(src_2)
\end{equation}

In practice, reproducing builds is still challenging and often fails~\cite{shi2021experience,drexel2024reproducible,randrianaina2024options,keshani2024aroma,malka2024reproducibility}. 
There are two basic approaches to address this issue. The first one is to control the environment sufficiently to avoid build variability. 
In the Java ecosystem, this is facilitated by reproducible central build specs~\footnote{\url{https://reproducible-builds.org/docs/jvm/}} that can be written by developers, or even be inferred by tools like \textit{macaron}~\cite{hassanshahi2025unlocking}.
There are some fundamental limitations to this approach, including non-determinism and the use of variability as a security feature in techniques like \textit{ASLR} (Address Space Layout Randomization)~\cite{lamb2021reproducible,shi2021experience}. 
The second approach is to move to \textit{explainable builds}~\cite{de2014challenges}. This accepts that full reproduction (i.e. the same binaries) can often not be achieved, and tools are needed to analyse differences, facilitating security assessment. Tools that can be used for this purpose include \textit{BuildDiff}~\cite{macho2017extracting}, \textit{RepLoc}~\cite{ren2018automated}, \textit{RepTrace}~\cite{ren2019root}, \textit{DiffoScope}~\cite{diffoscope} and \textit{JarDiff}~\cite{jardiff}.

A semi-formal approach to explainable builds is to weaken the condition of equality of packages to explainable equivalence~\cite{dietrich2025levels}. 
I.e. the conclusion in the build condition above is weakened to $build_1(src_1) \simeq build_2(src_2)$,  using some suitable equivalence relation $ \simeq$ between binaries.
Such an equivalence must under-approximate (undecidable) behavioural equivalence, and therefore comes with security guarantees in the sense that equivalent binaries will have the same behaviour. 
However, it can tolerate some variability, for instance caused by the use of different compilers or compiler versions. 
This therefore also reduces the level of trust required in up-stream components like compilers.  The effectiveness of this approach was evaluated in \cite{dietrich2025daleq}. Using \textit{daleq}, a datalog-based implementation of a binary equivalence relation, this was shown to be highly effective to establish the consistency of packages built by developers and released on Maven Central with packages built independently by Google's Assured Open Source (\gaoss) and Oracle's Build-From-Source (\bfs) projects.

Like Maven Central, \gaoss and \bfs also distribute source code with the binaries built. This is facilitated by the \textit{maven-source-plugin}, and the fact that publishing sources is a requirement to publish on Maven Central~\footnote{\url{https://central.sonatype.org/publish/publish-portal-maven/}}.  
I.e. builds usually produce an archive containing the sources used in builds alongside binaries.
In  \cite{dietrich2025daleq}, the prerequisite that sources should be same was tested. Surprisingly, sources often differed, but in most cases that could be attributed to differences caused by formatting or (legal) comments. Therefore, the approach in \cite{dietrich2025daleq} was to weaken the prerequisite to source equivalence tolerating those changes: 

\vspace{-0.3cm}
\begin{equation}
	src_1 \simeq_{src} src_2 \Rightarrow build_1(src_1)\simeq_{bin}build_2(src_2)
\end{equation}

Note that the notion of equivalence is different for source and binary equivalence. For source code, equivalence means that sources are the same once comments are removed and formatting in normalised. For byte code equivalence, this means that patterns of instructions are identified as behavioural equivalent. 

It appears that there are packages in the dataset used in~\cite{dietrich2025daleq} for which sources were not even equivalent. In the~\cite{dietrich2025daleq} study, those packages were excluded.  A possible explanation  is that the alternative builds failed to locate the correct commit, and those sources corresponded to a different commit. I.e. the developer build resulting in the artifact on Maven Central and the alternative build resulting in the \gaoss and/or \bfs artifact were based on different commits. 
This is a common problem when reproducing builds~\cite{fourne2023s,keshani2024aroma,hassanshahi2025unlocking}.
In this paper, we test this assumption by diffing those sources, and conduct a cause analysis.


Non-equivalent sources will usually also result in different binaries, so in this sense those issues are a root cause for reproducible builds failing.  
This study also provides insights into the origins of source code. 
It turns out that source code generation at build time is common, and provides challenges to build security, such as lack of reproducibility, non-determinism, and the use of third party build plugins that can increase the attack surface for software supply chain attacks on build pipelines.
As we will demonstrate, such plugins are widely used in Java, applications include annotation processing, the generation of type-safe API for resource access, and the generation of parsers for domain-specific languages (DSLs). 

We set out to answer the following two research questions:

\begin{enumerate}
	\item [RQ1] What are the causes of non-equivalent sources?  (Section \ref{sec:rq1})
	\item [RQ2] Are there sources that cannot be traced to source code in the repository?   (Section \ref{sec:rq2})
\end{enumerate}

We will also discuss possible approaches to improve security and provenance of build time generated code (Section \ref{sec:discussion}).

\section{Data Set}
\label{sec:dataset}

We have used the dataset from \cite{dietrich2025daleq}. This compares Maven artifacts built by open source developers and published on Maven Central with alternative builds by Google (\gaoss) and Oracle (\bfs), respectively.  
This dataset contains 2,714 pairs of separately built artifacts (jars) for the same GAV (group-artifact-version) coordinates, with 265,690 pairs of classes in bytecode format. 135,425  pairs  compare  \mvnc with \bfs built classes, and 130,265  pairs compare   \mvnc with \gaoss built classes.  In \cite{dietrich2025daleq} it was reported that 35,628 or those pairs are bitwise different. 
The focus of \cite{dietrich2025daleq} was to test whether building the same sources results in the same binary package, and to establish  a notion of equivalence between binary packages. 
In those experiments, the assumption that the two builds started from the same sources was tested using the notion of \textit{equivalence}, and if it failed, the respective records were excluded.
The notion of  source equivalence was introduced in order to take into account two modifications often made to sources when re-building packages: (1) comments are ignored to account for legal comments being inserted or removed and (2) formatting is ignored to account for reformatting, and issues such as OS-specific line separators. 

In this paper we set out to study those records excluded from \cite{dietrich2025daleq} due to non-equivalence of sources. 
The number of those records is relatively small (268), however, they effect many artifacts (85 artifacts, 28 if versions are ignored). 
An overview of those packages is provided in Table~\ref{table:dataset}. Those packages are all commodity libraries, widely used and backed by reputable projects and companies like \textit{Eclipse (Glassfish)}, \textit{Apache} and \textit{Google}. The last column lists package categories and rankings from \url{mvnrepository.com}~\footnote{The data was collected on February 2 2026}. Those rankings and categories are not available for all packages, in particular, they are missing for packages that are part of a coherent package set identified by a common group id. In this case, category and ranking is only provided for the ``main'' package within this group. 

\begin{table*}[!ht]
	\centering
	\footnotesize
	\caption{Dataset. Abbreviations: 	org.glassfish.jersey = \$ogj}
	\label{table:dataset}
	\begin{tabular}{|l|l|l|p{4cm}|}
		\hline
		package (GAV) & releases (GAs) & alt builder & ranking and category \\ \hline
		com.github.javaparser:javaparser-core & 3.24.4-3.25.8 (14 versions) & GAOSS & \#1 in Java Compilers/Parsers \\ \hline
		com.google.auto.service:auto-service & 1.0.1 & GAOSS & \#6 in Configuration Libraries ~ \\ \hline
		com.google.protobuf:protobuf-java & 3.25.1 & GAOSS & \#1 in Data Formats,  \#1 in Object Serialization~ \\ \hline
		com.google.truth:truth & 1.1.3,1.1.5,1.2.0 & GAOSS & \#2 in Assertion Libraries ~ \\ \hline
		io.netty:netty & 3.10.6.Final & GAOSS & \#1 in Network App Frameworks ~ \\ \hline
		io.rest-assured:rest-assured & 5.2.1 & GAOSS & \#14 in Testing Frameworks \& Tools~ \\ \hline
		io.undertow:undertow-core & 2.2.23.Final-2.3.6.Final (11 versions) & GAOSS & \#6 in Web Servers ~ \\ \hline
		io.undertow:undertow-servlet & 2.2.23.Final-2.3.6.Final (12 versions) & GAOSS & - ~ \\ \hline
		org.antlr:ST4 & 4.3.3 & GAOSS &  \#7 in Template Engines ~ \\ \hline
		org.apache.commons:commons-text & 1.11.0 & OBFS & \#1 in String Utilities~ \\ \hline
		org.apache.hadoop:hadoop-common & 3.3.5,3.3.6,3.4.0 & GAOSS & - ~ \\ \hline
		org.apache.hadoop:hadoop-hdfs & 3.3.5,3.3.6,3.4.0 & GAOSS & \#1 in Distributed File Systems ~ \\ \hline
		org.apache.zookeeper:zookeeper & 3.9.1 & OBFS & \#1 in Distributed Coordination ~ \\ \hline
		\$ogj.bundles:jaxrs-ri & 3.0.12 & OBFS & - ~ \\ \hline
		\$ogj.connectors:jersey-netty-connector & 3.0.12 & OBFS & - ~ \\ \hline
		\$ogj.containers:jersey-container-grizzly2-http & 3.0.12 & OBFS & - ~ \\ \hline
		\$ogj.containers:jersey-container-jdk-http & 3.0.12 & OBFS & - ~ \\ \hline
		\$ogj.containers:jersey-container-servlet & 3.0.12 & OBFS & - ~ \\ \hline
		\$ogj.containers:jersey-container-servlet-core & 3.0.12 & OBFS & - ~ \\ \hline
		\$ogj.containers:jersey-container-simple-http & 3.0.12 & OBFS & - ~ \\ \hline
		\$ogj.core:jersey-client & 2.37-3.0.12 (9 versions) & OBFS GAOSS & - ~ \\ \hline
		\$ogj.core:jersey-server & 2.37-3.0.12 (9 versions) & OBFS GAOSS & - ~ \\ \hline
		\$ogj.inject:jersey-hk2 & 3.0.12 & OBFS & - ~ \\ \hline
		\$ogj.media:jersey-media-jaxb & 3.0.12 & OBFS & - ~ \\ \hline
		\$ogj.media:jersey-media-sse & 3.0.12 & OBFS & - ~ \\ \hline
		org.hdrhistogram:HdrHistogram & 2.1.12 & GAOSS & \#6 in Application Metrics ~ \\ \hline
		org.hsqldb:hsqldb & 2.7.1,2.7.2 & GAOSS & \#2 in Embedded SQL Databases ~ \\ \hline
		org.mariadb.jdbc:mariadb-java-client & 3.4.0 & GAOSS & \#7 in JDBC Drivers and \#3 in MySQL Drivers~ \\ \hline
	\end{tabular}
\end{table*}
\section{RQ1: Causes of Non-Equivalent Sources}
\label{sec:rq1}

To answer RQ1, we have analysed the diffs generated for source code files distributed by both builds compared, and classified them.  An overview of the results is provided in Table~\ref{table:causes}. The first column lists packages identified by group and artifact ids (GA), the second column the number of releases for those packages identified by group and artifact ids and versions (GAV), the third column contains the number of unique classes (\textit{*.java} source code files) that are different across all releases, the forth column the alternative builder (the first builder is always the developer build on Maven Central (\mvnc)), the final column contains the diff category described in  more detail below.

\begin{table}[!h]
	\footnotesize
	\centering
	\caption{Cause analysis by package. Abbreviations: 	org.glassfish.jersey = \$ogj, cl. = number of different classes (.java files) across all versions}
	\label{table:causes}
	    \begin{tabular}{|l|l|l|}	\hline
			package (GA) & cl. & cause(s) \\ \hline
			com.github.javaparser:javaparser-core & 1 & codegen/meta \\ \hline
			com.google.auto.service:auto-service & 1 & inconsistentcommit \\ \hline
			com.google.protobuf:protobuf-java & 11 & codegen/proto \\ \hline
			com.google.truth:truth & 6 & codegen/@generated \\ \hline
			io.netty:netty & 1 & codegen/meta \\ \hline
			io.rest-assured:rest-assured & 11 & code/groovy \\ \hline
			io.undertow:undertow-core & 5 & codegen/@generated \\ \hline
			io.undertow:undertow-servlet & 2 & codegen/@generated \\ \hline
			org.antlr:ST4 & 1 & codegen/antlr \\ \hline
			org.apache.commons:commons-text & 1 & inconsistentcommit \\ \hline
			org.apache.hadoop:hadoop-common & 14 & codegen/proto , shading \\ \hline
			org.apache.hadoop:hadoop-hdfs & 12 & codegen/proto , shading \\ \hline
			org.apache.zookeeper:zookeeper & 2 & codegen/meta \\ \hline
			\$ogj.bundles:jaxrs-ri & 5 & codegen/istack \\ \hline
			\$ogj.connectors:jersey-netty-connector & 1 & codegen/istack \\ \hline
			\$ogj.containers:jersey-container-grizzly2-http & 1 & codegen/istack \\ \hline
			\$ogj.containers:jersey-container-jdk-http & 1 & codegen/istack \\ \hline
			\$ogj.containers:jersey-container-servlet & 1 & codegen/istack \\ \hline
			\$ogj.containers:jersey-container-servlet-core & 1 & codegen/istack \\ \hline
			\$ogj.containers:jersey-container-simple-http & 1 & codegen/istack \\ \hline
			\$ogj.core:jersey-client & 2 & codegen/istack \\ \hline
			\$ogj.core:jersey-server & 1 & codegen/istack \\ \hline
			\$ogj.inject:jersey-hk2 & 1 & codegen/istack \\ \hline
			\$ogj.media:jersey-media-jaxb & 1 & codegen/istack \\ \hline
			\$ogj.media:jersey-media-sse & 1 & codegen/istack \\ \hline
			org.hdrhistogram:HdrHistogram & 1 & codegen/meta \\ \hline
			org.hsqldb:hsqldb & 4 & codegen/meta \\ \hline
			org.mariadb.jdbc:mariadb-java-client & 3 & inconsistentcommit \\ \hline
	\end{tabular}
\end{table}

\subsection{Differences Caused by Generated Code}

Most differences are found in source code generated at build time. This can either mean that entire source code files are generated, or that existing source code is modified, e.g. by injecting additional fields or methods, or binding variables present in source code files (i.e. instantiating a source template).

\subsubsection{codegen/meta}
\label{sec:rq1:codegen:meta}

A common issue is that project and build metadata are made part of project classes, often as fields. 
E.g. \textit{io.netty:\-netty} encodes the project version in a field \textit{Version::ID} that is populated at build time. This differs between the \mvnc and the \gaoss builds (see Listing~\ref{diff-examples/meta1}).  

\begin{lstlisting}[caption={Diff caused by buildtime-generated metadata in io.netty:netty:3.10.6.Final, Version.java, \mvnc vs \gaoss}, label=diff-examples/meta1]
- public static final String ID = "3.10.6.Final-5f56a03";
+ public static final String ID = "3.10.6.Final-5f56a03bcb";
\end{lstlisting}

\textit{com.github.javaparser:javaparser-core} goes one step further, representing various build parameters as fields in the class \textit{JavaParserBuild}, leading to multiple differences across builds (see Listing~\ref{diff-examples/meta2}). 
A similar issue causes differences across builds of \textit{org.hdr\-histo\-gram:Hdr\-Histogram}.

\begin{lstlisting}[caption={Diff caused by buildtime-generated metadata in com.github.javaparser:javaparser-core:3.25.8, JavaParserBuild.java, \mvnc vs \gaoss}, label=diff-examples/meta2]
-    public static final String MAVEN_VERSION = "3.9.6";
+    public static final String MAVEN_VERSION = "3.8.6";
-    public static final String JAVA_VERSION ="1.8.0-292";
+    public static final String JAVA_VERSION ="1.8.0_342";
\end{lstlisting}

We also observed similar problems in \textit{org.apache.zookeeper:zoo\-keeper} (\mvnc vs \bfs). In \textit{VersionInfoMain::main}, differences are caused by version numbers printed to the console. The \bfs version of the class \textit{Info} contains the field \texttt{REVISION\_HASH="\${mvngit.\-commit.\-id}"}, suggesting the \bfs build had failed to bind a variable. 
Similar issues with failure to resolve a variable (\textit{\$REVISION\$}) cause differences in \textit{org.hsqldb:hsqldb} builds (\mvnc vs \gaoss).

\subsubsection{codegen/@Generated}
\label{sssec:rq1:generated}
This is a particular subcategory of \textit{codegen/meta} where the \textit{javax.annotation.Generated} or \textit{javax.\-anno\-tation.pro\-cessing.Generated} annotations are used to mark generated code. In the \textit{io.undertow} packages, the differences are caused by the \textit{date} property of those annotations that have build time timestamps as values. 
In  \textit{com.google.truth:truth:1.1.3}, the \gaoss built sources use the \textit{@Generated} annotation, whereas the \mvnc built sources only use a comment (see Listing~\ref{diff-examples/gen1}). This is caused by a different version and/or configuration of the annotation processor used for code generation. 

\begin{lstlisting}[caption={Diff caused by buildtime-generated metadata in com.google.truth:truth:1.2.0, AutoValue\_Actual\-Value\-Inference\_SubjectEntry.java, \mvnc vs \gaoss}, label=diff-examples/gen1]
-// Generated by com.google.auto.value.processor.AutoValueProcessor
+@Generated("com.google.auto.value.processor.AutoValueProcessor")
\end{lstlisting}

\subsubsection{codegen/istack}

\textit{istack-commons-maven-plugin} is  a Maven plugin that generates classes named \texttt{LocalizationMessages} at build time to access localization resources (.properties) through a type-safe API. 
Inspection of the respective source diffs reveals that a main difference is the order of methods and fields. At first sight this suggests non-determinism of the plugin. However, there are other differences. Notably, the  \mvnc versions contain a private parameterless constructor,  whereas the classes generated by the \bfs and \gaoss builds do not. This suggests that this is a plugin configuration and/or versioning issue. 

\subsubsection{codegen/proto}

Another source of generated code is the protobuffer compiler. The generated code differs across builds. 
An example of this is shown in Listing~\ref{diff-examples/protobuf}.  

\begin{lstlisting}[caption={Diff caused by the protobuf compiler in org.apache.hadoop:hadoop-hdfs:3.4.0, FsImageProto.java, \mvnc vs \gaoss}, label=diff-examples/protobuf]
   public interface FileSummaryOrBuilder extends
       // @@protoc_insertion_point(interface_extends:hadoop.hdfs.fsimage.FileSummary)
-      org.apache.hadoop.thirdparty.protobuf.MessageOrBuilder {
+      com.google.protobuf.MessageOrBuilder {
\end{lstlisting}

In this particular case different classes are referenced, we will discuss this further in Section~\ref{ssec:rq1:shading}.

\subsubsection{codegen/antlr}
\label{sssec:rq1:antlr}

Domain-specific languages are widely used in modern software engineering. Grammar libraries such as \textit{antlr} support the generation of parsers at build time through build plugins, therefore facilitating the consistency between grammar and parser. 
The string templating package \textit{org.antlr:ST4:4.3.3} uses this approach, leading to different sources across the \mvnc and \gaoss builds, respectively. Inspecting the diff(s) suggests that the differences are all related to the order of elements, see Listing~\ref{diff-examples/antlr} for an example. Note that the order of checks is potentially behaviour-changing unless the \texttt{hasNext()} methods are side effect-free.

\begin{lstlisting}[caption={Diff caused by the antlr parser generator in org.antlr:ST4:4.3.3, STParser.java, \mvnc vs \gaoss}, label=diff-examples/antlr]
-	while ( stream_t2.hasNext()||stream_ELSEIF.hasNext()||stream_c2.hasNext() ) {
+  while ( stream_c2.hasNext()||stream_t2.hasNext()||stream_ELSEIF.hasNext() ) {
\end{lstlisting}

This suggests that the \textit{antlr} parser generator might be non-deterministic. 
We have confirmed non-determinism for the parser code generated by \textit{antrl} for \textit{org.antlr:ST4:4.3.3}. For this purpose, we built the project ten times, and diffed the \textit{STParser.java }sources generated in \textit{target/generated-sources/}. This reveals differences similar to the ones shown in Listing~\ref{diff-examples/antlr}. We have opened an issue for this in the ST4 GitHub project~\footnote{\url{https://github.com/antlr/stringtemplate4/issues/325}}.

\subsubsection{codegen/groovy}

For \textit{io.rest-assured:rest-assured:5.2.1}, differences are caused by differences in the (sometimes redundant) use of qualified type names. Closer inspection reveals that those files are stubs generated by the groovy compiler as part of the build (via the \textit{gmaven-plugin} plugin). The structure of the diffs suggests that the issue is caused by configuration and/or version of the plugin and the underlying groovy compiler. 
A particular idiosyncrasy of the groovy compiler is that those sources do not directly correspond to bytecode. They are temporary constructs used to type-check the bytecode that is being generated by groovy compiler.
 
\subsection{Shading}
\label{ssec:rq1:shading}

Shading is a technique applied in order to rename packages, usually in order to avoid classpath conflicts. This is often done dynamically at runtime by using Maven plugins such as \textit{maven-shade-plugin}. 
Shading causes many challenges for software composition analysis if dependencies on cloned or shaded components are not declared~\cite{dietrich2023security}.

For the \textit{hadoop} packages, shading is used to support protocol buffers. 
I.e., instead of using packages (classes) from \textit{com.google.\-proto\-buf:protobuf-java}, packages shading those such as \textit{org.apache.ha\-doop.thirdparty:hadoop-shaded-protobuf\_3\_7} are used~\footnote{For instance, see \url{https://central.sonatype.com/artifact/org.apache.hadoop/hadoop-common/3.3.6}}.  
The \gaoss builds do not use the shaded packages (names starting with \textit{org.\-apache.\-hadoop.thirdparty.protobuf}) but instead uses the original protobuf packages (names starting with \textit{com.google.protobuf}). This can lead to subtle behavioural differences. An example is shown in Listing~\ref{diff-examples/protobuf}.

\subsection{Inconsistent Commits}
\label{ssec:rq1:commits}

For several packages we have identified changes that suggest that the respective builds have used different commits. We have used \textit{git log} and \textit{git bisect}  to identify those commits. An overview of those packages is given in Table~\ref{table:commits}. We also include the git tags corresponding to the respective releases. 

The commit for  \textit{com.google.auto.service:auto-service:1.0.1} suggest that the rebuild used changes made after the original release. Those changes are related to the refactoring (moving) of a static method being used. 

For \textit{org.apache.commons:commons-text:1.11.0}, the release tag refers to the commit that corresponds to the diffs encountered. However, those changes are missing in the \mvnc build. A possible explanation is that the \mvnc build and release was made first, and the tag was either created shortly after pointing to a subsequent commit, or moved.  However,  \bfs has identified the commit using this tag, making the builds inconsistent. 

Similarly, for \textit{org.mariadb.jdbc:mariadb-java-client:3.4.0}, the \mvnc build contains code for a later commit, not present in the commit referenced by the release tag. The \gaoss build however relies on the release tag.

\begin{table}[h!]
	\footnotesize
	\begin{tabular}{|r|lll|}
		\hline
		package (GAV)      & diff & release & tag   \\ \hline 
		\textit{..:auto-service:1.0.1} & 88d2744    & 4695896    & auto-service-1.0.1   \\
	    \textit{..:commons-text:1.11.0} &     d5e8a73      &    d5e8a73 &    rel/commons-text-1.11.0 \\
	    \textit{..:mariadb-java-client:3.4.0} & 7b169ce   & c19f608 &  3.4.0 \\
		\hline 
	\end{tabular}
	\caption{Different commits corresponding to diffs between .java files vs release commits}
	\label{table:commits}
\end{table}

\section{RQ2: Consistency of Source Code with Sources in Repository}
\label{sec:rq2}

To answer the second research question, we used the records described in Section~\ref{sec:dataset}, and located the repositories and commits. 
We used \textit{macaron (version 0.20.0)} for this purpose.  We located the repositories for all packages except the following.  
For \textit{org.apache.hadoop:\-hadoop-common} and \textit{org.apache.hadoop:hadoop-hdfs} (versions \textit{3.3.5},  \textit{3.3.6} and \textit{3.4.0}), \textit{macaron} failed due to missing source repository information in the project poms. 
We located the respective commits manually based on tags corresponding to the respective versions. 
For \textit{org.hsqldb:hsqldb} versions \textit{2.7.1} and \textit{2.7.2}, we manually located the sources on \textit{SourceForce}, using SVN, and matched the versions to SVN tags.
  
 We then cloned those repositories, checked out those commits, and compared the sources distributed with the alternative builds with the sources found in commits. 
 To identify \textit{valid} java sources in the commits, we did not only search the project folders for files with the \textit{.java} extension, but also checked for compliance with the Java language specification~\cite{JLS7Spec}. We used the \textit{javaparser} library~\cite{javaparser} for this purpose. 
 We also extracted package and class names from \textit{.java} files using \textit{javaparser}. 
 This way we created lists of qualified names of Java classes defined as top-level classes in the respective \textit{.java} files, and then compared those lists with the sources distributed with packages.
  
  
\begin{table}[!ht]
	\centering
	\caption{Source code files distributed with alternative builds not found in the source repository. Abbreviations: 	org.glassfish.jersey = \$ogj}
	\label{table:consistency}
	\footnotesize
		    \begin{tabular}{|l|l|} \hline
				package (GAV) & classes missing in commit\\ \hline
				com.google.protobuf:protobuf-java & 11 \\ \hline
				com.google.truth:truth & 6 \\ \hline
				io.netty:netty & 1 \\ \hline
				io.rest-assured:rest-assured & 11 \\ \hline
				io.undertow:undertow-core & 5 \\ \hline
				io.undertow:undertow-servlet & 2 \\ \hline
				org.antlr:ST4 & 1 \\ \hline
				org.apache.hadoop:hadoop-common & 14 \\ \hline
				org.apache.hadoop:hadoop-hdfs & 12 \\ \hline
				org.apache.zookeeper:zookeeper & 1 \\ \hline
				\$ogj.bundles:jaxrs-ri & 5 \\ \hline
				\$ogj.connectors:jersey-netty-connector & 1 \\ \hline
				\$ogj.containers:jersey-container-grizzly2-http & 1 \\ \hline
				\$ogj.containers:jersey-container-jdk-http & 1 \\ \hline
				\$ogj.containers:jersey-container-servlet & 1 \\ \hline
				\$ogj.containers:jersey-container-servlet-core & 1 \\ \hline
				\$ogj.containers:jersey-container-simple-http & 1 \\ \hline
				\$ogj.core:jersey-client & 2 \\ \hline
				\$ogj.core:jersey-server & 1 \\ \hline
				\$ogj.inject:jersey-hk2 & 1 \\ \hline
				\$ogj.media:jersey-media-jaxb & 1 \\ \hline
				\$ogj.media:jersey-media-sse & 1 \\ \hline
				org.hdrhistogram:HdrHistogram & 1 \\ \hline
		\end{tabular}
\end{table}

The result of this comparison is summarised in Table~\ref{table:consistency}. This shows the packages, and the total number of java source files missing in commits. 
If a source with a unique name is missing in more than one release, it is only counted once.
We found that for 23/28 packages, at least one java source file is missing. 

Comparing Tables~\ref{table:causes} and \ref{table:commits} suggests that for those package differences in sources distributed correspond to classes missing in the repository, i.e. classes that have been generated. 
There is one exception though. For  \textit{org.apache.zookeeper:zookeeper} the number of classes don't match (2 different classes, but only 1 missing in the repository), one class is different but is not missing in the repository. 
The two classes that are different across builds are \textit{Info} and \textit{VersionInfoMain}, both in the \textit{org.apache.zookeeper.version} package. 
Listings~\ref{diff-examples/zookeeperInfo.java} and \ref{diff-examples/zookeeperInfoMain.java} respectively show the partial source code of those classes. Both are templates with variables instantiated at build time. 
The only difference is that in \textit{VersionInfoMain}, the variable is part of a string literal, whereas in \textit{Info}, it is used to initialise a field. I.e. \textit{VersionInfoMain} is still a valid Java file, whereas \textit{Info} is not. This explain the difference.

\begin{lstlisting}[caption={(Partial) org.apache.zookeeper:zookeeper class Info.java from the GitHub repository}, label=diff-examples/zookeeperInfo.java]
public interface Info {
    int MAJOR=${parsedVersion.majorVersion};
    int MINOR=${parsedVersion.minorVersion};
\end{lstlisting}

\begin{lstlisting}[caption={(Partial) org.apache.zookeeper:zookeeper class VersionInfoMain.java from the GitHub repository}, label=diff-examples/zookeeperInfoMain.java]
public class VersionInfoMain implements org.apache.zookeeper.version.Info {
    public static void main(String[] args) {
        System.out.println("Apache ZooKeeper, version ${project.version} ${build.time}");
    }
}
\end{lstlisting}

There are five packages for which no differences were detected, this is the difference between the 28 packages in Table~\ref{table:causes} and the 23 packages in Table~\ref{table:consistency}. For three packages \textit{com.google.auto.service:auto-service} , \textit{org.apache.\-commons:commons-text}  and \textit{org.maria\-db.jdbc:\-maria\-db-java-client}, the differences are caused by a different commit being used by the alternative build (see Section~\ref{ssec:rq1:commits}), but the class names are same for both commits. The two remaining packages \textit{com.github.java\-parser:javaparser-core}  and \textit{org.hsqldb:hsqldb} both use templating, but variables occur only in string literals and comments. Therefore, the respective \textit{.java }files are still valid Java sources.

\section{Discussion}
\label{sec:discussion}

\subsection{Threats to Validity}

We have only compared \textit{.java} files, no kotlin, groovy, scala etc source that might have been present in the respective projects.
We might therefore be slightly under-report the issues around generated code. 
However, to the best of our knowledge code generators typically emit java source code.

We cannot fully guarantee that the consistency of commits as the sources distributed with binaries do not contain all configuration and resource files that are necessary to do so. 
Therefore, there might be issues in \ref{ssec:rq1:commits} that are under-reported. 
I.e. changes in the configuration of a source-generating plugin could be caused by changes to a configuration file made in a commit. 

\subsection{Challenges}

Binaries generated by different compilers or compiler versions can vary widely, for instance, due to changes in the compilation strategy and/or optimisations~\cite{jep181,jep280}. 
What is more, compilers cannot be assumed to be deterministic~\cite{xiong2022towards}. 
In Java, compiler non-determinism is seen as not desirable and if discovered is treated as a bug~\footnote{Recent OpenJDK issues include: JDK-8264306, JDK-8072753, JDK-8076031 and JDK-8295024 (URL: \textit{https://bugs.openjdk.org/browse/$<$issue-id$>$}).}. 
However, code generators are  another source of variability and non-determinism. 
We have demonstrated in Section~\ref{sssec:rq1:antlr} that the \textit{antlr} parser generator is non-deterministic, therefore making the entire build non-deterministic. 
Similarly, any code generator using the \texttt{@Generated} annotation with a build timestamp is inherently non-deterministic due to the use of timestamps, see Section~\ref{sssec:rq1:generated}.

Traditionally, the compiler was seen as the weak point in build security~\cite{thompson1984reflections}. 
This study suggests that build plugins generating code are a major contributor to build variability and security. 
In particular, a compromised plugin generating code at build time can generate malicious code. 
Therefore, those plugins deserve more attention.

\subsection{Suggestions}

We found that the main issue causing non-equivalence of sources across builds is build-time code generation. 
In our analysis, we have manually associated source code with those code generators. 
It would be highly desirable to develop techniques that facilitate automated reasoning by security tools. 
This could then be used to locate root causes of non-equivalence, and reason about them. 
For instance, if there is known non-determinism in a given code generator, supply chain security tools could take this into account when comparing and assessing different builds.  
In particular, the non-equivalence of generated sources can be treated differently from non-equivalence due to the use of different commits. 
Security tools could also be configured to reject packages containing binaries derived from sources generated by black listed code-generators.

A good starting point for making generated code explicit is the \texttt{@Generated} annotation.  
This used to be in the\textit{ javax.annotation} package, and has been moved in Java 9 to \textit{javax.annotation.processing}. 
There are a number of limitations with the current version: 
\begin{enumerate*}[label=(\arabic*)]
	\item The retention of the annotation is set to \textit{source}, therefore limiting the code analyses for which it can be used. Many static analysis tools use bytecode as input. 
	\item The annotation is defined in the \textit{java.compiler} module, therefore imposing additional dependencies~\footnote{See for instance \url{https://github.com/eclipse-ee4j/jaxb-istack-commons/blob/fe7020a077a283f6c221302426250258a8dc1b22/istack-commons/maven-plugin/src/main/java/com/sun/istack/maven/QuickGenMojo.java\#L225}}. This is also an issue with similar annotations in third party libraries, such as \textit{jakarta.annotation.Generated}.
	\item The \textit{date} field is a source of build non-determinism (see Section~\ref{sssec:rq1:generated}). 
	\item The \textit{value} field is the fully qualified name of the code generator. The semantics is vague~\footnote{\url{https://docs.oracle.com/en/java/javase/24/docs/api/java.compiler/javax/annotation/processing/Generated.html}}, but the way this is used suggests that this is the fully qualified name of a Java class.
\end{enumerate*}

To make it more useful to address use cases in software supply chain security, a better \textit{@Generated} annotation should therefore satisfy the following properties:
\begin{enumerate*}[label=(\arabic*)]
	\item Have a retention policy runtime to facilitate a wide range of analyses, including bytecode-based analyses.
	\item Be defined in \textit{java.base}, avoiding dependencies on additional modules and third party libraries.
	\item Do not use a date or timestamp field, or set a corresponding property to enable the reproducible builds mode~\footnote{\url{https://maven.apache.org/guides/mini/guide-reproducible-builds.html}}.
	\item For the \textit{value} property, develop and use a naming scheme for code generators. This could be Java classes, Maven plugins, external tools like DSL compilers, etc. Names should include versions, and configurations that have an impact on the code being produced.  Schema like \textit{purl}~\footnote{\url{https://github.com/package-url/purl-spec}} are good examples of what this could look like. 
\end{enumerate*}

Notably, the use of such an annotation could be enforced at build time, for example, by utilizing a Maven plugin that compares project sources with those provided as compiler input. 
Alternatively, information about code generators, as well as the tools and configurations used during the build, can be captured separately in an extended version of the source SLSA provenance~\cite{source-prov}.
Part of this is to lock the versions of plugins, extending approaches to lock dependencies in order to facilitate rebuilds~\cite{schmid2025maven}.

\section{Related Work}
\label{sec:relatedwork}

Benedetti et al. studied build reproducibility of open source packages across several package managers including Maven~\cite{benedetti2025empirical}. 
They found that the reproducibility of Maven packages remains challenging. 
Sharma et al. categorized causes of reproducible builds failing, and propose a tool \textit{Chains-Rebuild} to improve success rates~\cite{sharma2025canonicalization}.

Several author studied the particular challenge of how to locate the correct commit to rebuild, including \textit{aroma} by Keshani et al.~\cite{keshani2024aroma} and \textit{macaron} by Hassanshahi et al. \cite{hassanshahi2023macaron,hassanshahi2025unlocking}. This is an issue we have encountered and discussed in Section~\ref{ssec:rq1:commits}. We have used  \textit{macaron} for our experiments.
\textit{macaron} outperforms \textit{aroma} in locating the correct commits~\cite{hassanshahi2025unlocking}

Several authors tried to address other aspects that can lead to non-reproducibility. Notably Xiong et al. study issues preventing reproducibility of Java-based systems such as non-determinism and suggest tools to address those by normalising bytecode~\cite{xiong2022towards}. 
Schott et al. propose \textit{JNorm}~\cite{schott2024JNorm}, a tool to normalise bytecode based on similarity. 
Dietrich et al. provide a conceptual framework for bytecode similarity~\cite{dietrich2025levels} inspired by types of code clones~\cite{bellon2007comparison} and propose the \textit{daleq} tool~\cite{dietrich2025daleq} to provide an explainable equivalence relation for automated explainable builds, acknowledging that full reproducible builds are not always achievable.

Schmid et al. worked on approaches to pin dependencies to facilitate builds~\cite{schmid2025maven}. This lock file can also include Maven plugins, therefore providing provenance similar to what we suggest in Section~\ref{sec:discussion}.  We note however that if the Maven plugin is just a wrapper for some other separately installed tool (like a DSL compiler), then this approach is not sufficient. 

\section{Conclusion}
\label{sec:conclusion}

In this paper, we analysed instances of non-equivalent Java sources that occur in Maven packages when re-built at scale by Google and Oracle.
This represents a significant issue, as it affects several widely used packages in the Java ecosystem. Our analysis revealed that, in most cases, these discrepancies stem from source files generated at build time by various build plugins.
Such plugins contribute to differences in the produced binaries, can render builds non-deterministic, and consequently increase the attack surface for software supply chain attacks.
To address this challenge, we have proposed a technique aimed at improving the provenance and traceability of code generated by these plugins.

\bibliographystyle{plain}
\bibliography{references}

\end{document}